\documentclass[12pt]{article}
\usepackage{epsfig}
\newcommand{\beq}{\begin{equation}}
\newcommand{\eeq}{\end{equation}}
\newcommand{\beqs}{\begin{eqnarray}}
\newcommand{\eeqs}{\end{eqnarray}}
\newcommand{\ws}{W_{\mathrm{susy}}}
\newcommand{\wir}{W_{\mathrm{ir}}}
\newcommand{\phiir}{\phi_{\mathrm{ir}}}
\newcommand{\phis}{\phi_{\mathrm{susy}}}
\newcommand{\op}{{\cal O}_\phi}
\begin{document}
\begin{titlepage}
\begin{flushleft}
       \hfill                      {\tt hep-th/0003151}\\
       \hfill                       SISSA ref. 26/2000/EP\\
       \hfill                       G\"oteborg ITP Preprint\\
\end{flushleft}
\vspace*{1.8mm}
\begin{center}
{\bf\large A Study of Holographic Renormalization}\\
{\bf\large Group Flows in $d=6$ and $d=3$}\\
\vspace*{6mm}
{\large Vanicson L. Campos$^a$\footnote{\tt vanicson@fy.chalmers.se},
Gabriele Ferretti$^a$\footnote{\tt ferretti@fy.chalmers.se},
Henric Larsson$^a$\footnote{\tt solo@fy.chalmers.se},  
Dario Martelli$^b$\footnote{\tt dmartell@sissa.it}, 
Bengt E.W. Nilsson$^a$\footnote{\tt tfebn@fy.chalmers.se}}\\
\vspace*{3mm}
{\it $^a$ Institute of Theoretical  Physics\\
Chalmers University of Technology\\
412 96 G\"{o}teborg, Sweden}\\
\smallskip{\rm and\\}\smallskip
{\it $^b$ SISSA, Via Beirut 2 Trieste 340 14\\
and INFN, Sezione di Trieste, Italy\\}

\vspace*{5mm}
\end{center}
\begin{abstract}
We present an explicit study of the holographic renormalization group (RG) 
in six dimensions using minimal gauged supergravity.
By perturbing the theory with the addition of a relevant operator of 
dimension four one flows to a non-supersymmetric conformal 
fixed point. There are also solutions describing
non-conformal vacua of the same theory obtained
by giving an expectation value to the operator. One such vacuum is
supersymmetric and is obtained by using the true superpotential of the theory.
We discuss the physical acceptability of these vacua by applying the
criteria recently given by Gubser for the four dimensional case and
find that those criteria give a clear physical picture in the six
dimensional case as well. We use this example to comment 
on the role of the Hamilton-Jacobi equations in implementing the RG.
We conclude with some remarks on $AdS_4$ and the status of three 
dimensional superconformal theories from squashed solutions of M-theory.
\end{abstract}
\end{titlepage}

\section{Introduction and Summary}
The study of the AdS/CFT correspondence has been the major occupation of 
the {\tt hep-th}-community since the initial breakthroughs 
of~\cite{thethree}. Amongst the many aspects of the correspondence, one 
of the most intriguing is the possibility of formulating the field 
theoretical RG flow 
in terms of the classical dynamics of the gravitational theory in the 
bulk~\cite{rg,gira}. 

In studying the RG flow induced by certain operators 
from an ultra-violet (UV) fixed point,
one needs to have a dictionary associating each 
operator to the appropriate field in the bulk. This requires 
resolving some possible ambiguity~\cite{plusminus} arising 
for a specific range of bulk masses near the 
stability bound~\cite{bound}, which amounts to making a {\it choice} 
between two theories, both in principle described by the same bulk 
fields. Usually, only one such theory is supersymmetric and the allowed 
values for the conformal weight $\Delta$ can be read off from the 
representations of supersymmetry.
After the association is made, one still needs to determine whether 
one is deforming the UV Lagrangian by adding the operator itself
or simply going to a different vacuum where the operator acquires
a non-zero vacuum expectation value (vev).
(For a clear exposition of this point see~\cite{vevdef}). 
In view of this physical interpretation, not 
all the solutions to the gravity theory are acceptable,
for instance one should rule out flows in which a positive
definite operator acquires a negative vev. 
We will shortly discuss these conditions.

In a related development, it was noticed in~\cite{hj} that 
a formulation of gravity 
in first order Hamiltonian formalism provides further insights into
the RG 
\footnote{That the RG equations can be given a Hamiltonian structure was
noticed some years ago ~\cite{dolan}.}.
The double way of writing the equations -- one as a second order 
Lagrangian system supplemented by the zero energy constraint and the 
other as a first order system written in terms of a 
``superpotential'' -- is the origin of some confusion 
in the implementation of the holographic RG.
The name superpotential in this context is 
somewhat of a misnomer because, whereas the Hamilton-Jacobi equations 
have a continuum set of solutions parameterized by the constants of 
motion, only one such solution can be regarded as the superpotential 
arising in a supersymmetric theory of gravity. 
We shall reserve the name {\it 
superpotential} for the truly supersymmetric one and call all the  
solutions to the Hamilton-Jacobi equation {\it generating functions}.

In many circumstances, the few particular generating functions that can 
be found explicitly are precisely those that can be thought of as true 
superpotentials. 
If we are interested in flows between
two fixed points of which only one (typically the UV one) appears as an 
extremum of the known superpotential, we cannot use the first order 
equations for this purpose and must revert to the Lagrangian system. In 
this case, it is impossible to obtain an analytical solution 
and it is impossible to resolve the vev/deformation ambiguity by 
asymptotically expanding it near the UV fixed point.
How can one decide then which of these two cases is realized?
Also, does the fact that the known generating function cannot be used to 
connect two fixed points mean that it is useless in 
connection with the RG flow? Or, if it can be used, how can the same 
boundary operator induce different flows?

These questions have been addressed in above cited literature and 
a coherent picture has emerged. So far, most 
of the attention has been focused on the case of various gauged 
supergravities in $d=5$, which are, of course, the ones most relevant to four 
dimensional field theories.
We will address and solve these problems very explicitly for a 
particularly simple example -- ${\cal N}=1$ $d=7$ gauged 
supergravity~\cite{sugra7}. This example has all the features we want to 
study and its simple field content allows for a clear-cut solution\footnote
{The case of ${\cal N}=1$ $d=7$ gauged supergravity has been recently 
discussed in ~\cite{sugra7new}, where
some comments in the direction of the 
results of this paper have been made.}. 
By presenting a thorough and 
explicit solution of the problem, we hope to contribute in clarifying some 
points that might have remained obscure from the previous discussions.

The status of ${\cal N}=1$ $d=7$ gauged supergravity can be summarized 
as follows: First of all, there is only one scalar field $\phi$ in the 
bulk and its potential has two extrema (see figure~\ref{figpot}), a maximum at 
$\phi=0$ corresponding to a supersymmetric UV theory and a minimum 
at\footnote{In 
appropriate units to be specified later.} $\phi=-\log 2 /\sqrt{5}$, 
corresponding to a non-supersymmetric but nevertheless stable IR theory. 
The ``tachyonic'' excitation near the UV 
point has a mass $m$ given, in units of the AdS radius $r$, by $m^2r^2 = 
-8$. The boundary operator corresponding to $\phi$ is 
$\op={\Phi}^2$, where $\Phi$ is a scalar
in the tensor multiplet of the $d=6$ CFT
or, better, its still unknown non-Abelian generalization.
The conformal dimension of $\op$ is $\Delta = 4$. 
The other possibility ($\Delta=2$) 
is ruled out by looking at the table 1 of~\cite{multiplet1} for the
multiplets of extended  (${\cal N}=2$) supersymmetry and 
figure 2 in~\cite{multiplet2}. In fact, $\Delta=2$ corresponds to the
singleton field $\Phi$ itself.

{\it Deforming} the UV theory ($\phi=0$) 
by the addition of $\int\phi\op$ to the 
fixed point Lagrangian induces an RG flow that ends 
at the non-supersymmetric IR conformal fixed point. In this case the 
generating function cannot be obtained explicitly but 
it can be computed numerically and shown to have the correct 
behavior at both ends -- it corresponds to a particular one among
the 1-parameter family of solutions of the Hamilton-Jacobi equation.
This is the {\it only} solution in which the field $\phi$ is allowed
to acquire negative values. All other such solutions correspond to 
{\it negative} vev's for $\op$ and should be ruled out, in tune with 
the fact that, when evaluated
on solutions that are running away to $-\infty$, the potential is not
bounded from above and the metric singularity at the runaway point is
that of a naked time-like nature~\cite{gubser}.

It turns out that the only solution to the Hamilton-Jacobi equation that has an
extremum and is analytic
at $\phi=0$ is the superpotential. The superpotential can be used to
study new supersymmetric vacua of the theory, for which 
$\langle\op\rangle > 0$, by studying runaway solutions in which 
$\phi\to +\infty$. There is also a continuum set of solutions,
still with $\phi\to +\infty$,
describing what we believe are consistent non-supersymmetric vacua.

Towards the end of the paper
we will turn to the more complicated case of compactification of
$d=11$ supergravity~\cite{sugra11} on ``squashed'' manifolds ($\tilde S^7$ 
and $\tilde N(1,1)$) and comment on some particular features of the 
models, complementing the discussion in~\cite{ar}.

In~\cite{ar}, the interesting question was raised of whether there exist
trajectories connecting the squashed solutions with the 
corresponding unsquashed manifolds. The situation is similar to the
well-studied case of gauged $d=5$ supergravity, but there is a
crucial difference: In $d=5$ supergravity, the analog particle rolls
from a saddle point to a minimum of the (inverted) potential, 
whereas here it should roll from a maximum to a saddle point, clearly
a more unstable situation. If the RG
equations where truly first order, one could argue from general
theorems that there must still be a critical line connecting the
points. However, the equations expressed in terms of the potential are
second order and there is no guarantee that such a solution will
survive. In fact, we have reasons to believe that such a flow does
not exist, although more work is required to fully establish or 
refute this belief.

As far as the squashed solutions are 
concerned, one is able to find an explicit solution\footnote{The
explicit generating function has no fixed
point at the unsquashed vacuum (or else a solution connecting the two
would exist).} to the 
Hamilton-Jacobi equation. It turns out that this corresponds to giving
a vev to the squashing operator, thus breaking conformal invariance.
Given the simple form of the generating function and the collected 
experience with similar models, it is tempting to conjecture that such a 
solution is in fact supersymmetric, although here we are working beyond the 
gauged supergravity truncation and considering fields from the
higher levels of the Kaluza-Klein spectrum. 

\section{${\cal N}=1$ $d=7$ gauged supergravity}
\begin{figure}
\begin{center}
\epsfig{file=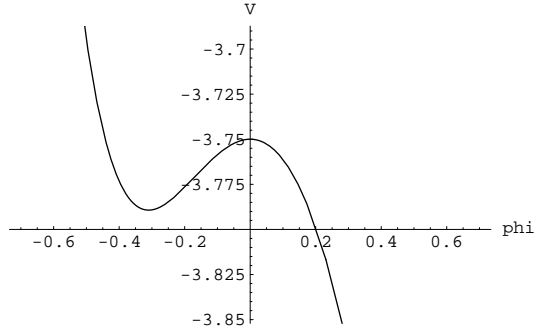,width=7 cm}
\caption{The potential $V(\phi)$ plotted against $\phi$.
The point $\phi=0$ represents the UV theory and the point 
$\phi=-\log 2/\sqrt{5}\approx -0.309985$ the IR theory.}\label{figpot}
\end{center}
\end{figure}

The field content, Lagrangian and supersymmetry transformations for 
${\cal N}=1$ $d=7$ gauged supergravity can be found in~\cite{sugra7}. For 
our purposes, we set all fields to zero except for the metric and the scalar 
$\phi$. The action is
\beq
        S=\int d^7x \; \sqrt{g}\left(\frac{1}{2} R - \frac{1}{2} 
        (\partial\phi)^2 - V(\phi)\right).  \label{act}
\eeq
The scalar potential is chosen to be\footnote{In its full generality, the 
potential depends on two arbitrary constants $h$ and $g$ and it displays
two minima as long as $h/g>0$ 
(c.f.r.~\cite{sugra7,ar}). One combination of  $h$ and $g$ is eliminated by 
shifting $\phi$ and the remaining one is an irrelevant overall 
multiplicative constant in front of the potential.}
\beq
        V(\phi) = \frac{1}{4} e^{-8\phi/\sqrt{5}} -
        2 e^{-3\phi/\sqrt{5}} - 2 e^{+2\phi/\sqrt{5}}.
\eeq
A plot of $V(\phi)$ is shown in figure~\ref{figpot}. There is a 
supersymmetric  UV fixed point at $\phi=0$ and a stable non-supersymmetric IR 
one at $\phi=-\log 2/\sqrt{5}$.
The Lagrangian equations of motion following from (\ref{act}), with the 
standard domain-wall ansatz
\beq
     ds^2=dy^2 + e^{2A(y)} \eta_{\mu\nu}dx^\mu dx^\nu ,
     \quad\hbox{and}\quad \phi = \phi(y) \label{ansatz}
\eeq
can be derived by the following action for a mechanical system
\beq
        S=\int dy \; e^{6A}\left(15 \dot A^2 - \frac{1}{2} 
        \dot\phi^2 - V(\phi)\right)~.  \label{actans}
\eeq
When supplemented by the zero energy constraint, the equations read
\footnote{The primes denote the derivative with 
respect to $\phi$ and the dots the derivative with respect to $y$.} 
\beqs
        \ddot{\phi} + 6 \dot A \dot\phi &=& 
       V^\prime(\phi) \label{lagrange}\\
       5 \ddot{A}+15\dot{A}^2+\frac{1}{2}\dot{\phi}^2 &=&
       -V(\phi)\label{lagrange2}\\
       15 \dot A^2 - \frac{1}{2} \dot\phi^2 &=& - V(\phi)~. 
        \label{constraint}
\eeqs
Equation (\ref{lagrange2}) can be easily shown to follow from  
(\ref{lagrange}) and (\ref{constraint}).

Equivalently, one can consider the equation for Hamilton's 
characteristic function 
$F(A,\phi, c)$ generating the canonical transformations to the cyclic 
coordinates\footnote{There is only one constant, $c$, because the other 
conjugate variable is set to zero by (\ref{constraint}).} 
\beq
        \frac{1}{60}\left(\frac{\partial F}{\partial A}\right)^2 - 
        \frac{1}{2} \left(\frac{\partial F}{\partial \phi}\right)^2 
        + e^{12A}V = 0~.
      \label{gene}
\eeq
By substituting the ansatz $F(A,\phi, c) = e^{6 A} W(\phi, c)$ into
(\ref{gene}) the equation becomes the same as the defining equation for 
the superpotential. Altogether, expressing the canonical transformation 
in terms of $W$ we end up with the first order system of 
Hamilton-Jacobi equations
\beqs
        \dot\phi &=& W^\prime \label{hj1} \\
        \dot A &=&  -\frac{1}{5} W \label{hj2}\\
         V &=& \frac{1}{2} {W^\prime}^2 - \frac{3}{5}W^2~. \label{sp}
\eeqs
Equation (\ref{sp}) is obeyed by the superpotential of the theory 
but it also admits a continuum of solutions, parameterized by $c$, that 
have nothing to do with supersymmetry. 
If one wants to recover all the solutions to the Lagrangian 
equations this way, one needs to consider all possible solutions to 
(\ref{sp}).

\begin{figure}
\begin{center}
\epsfig{file=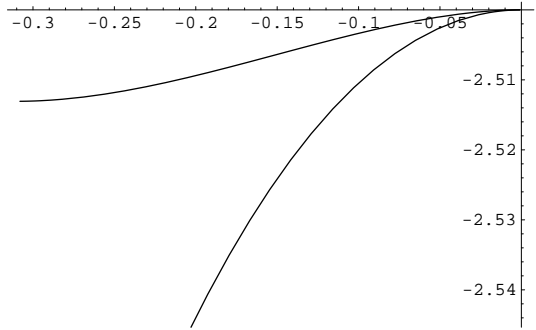,width=7 cm}
\caption{Comparison of the two generating functions 
$\wir$ and $\ws$ between the IR and UV fixed points. 
Note that $\wir$ has a second extremum at the IR
fixed point, whereas $\ws$ does not.}
\label{figagainst}
\end{center}
\end{figure}

\begin{figure}
\begin{center}
\epsfig{file=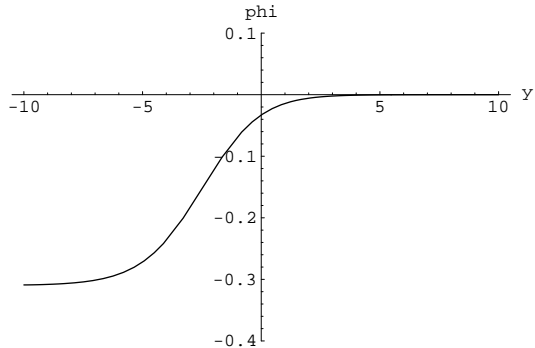,width=7 cm}
\caption{The solution $\phiir$ connecting the two fixed points plotted
against the scale factor $y$.}
\label{figconnect}
\end{center}
\end{figure}

Particularly confusing is the fact that there are different solutions to
(\ref{sp}) that have an extremum at $\phi=0$.
One solution, $\ws$,
can be easily found by inspection and identified with the superpotential
\footnote{This is defined up to an overall unimportant sign.}:
\beq
     \ws = -2 e^{\phi/\sqrt{5}} - \frac{1}{2} e^{-4\phi/\sqrt{5}}~. 
     \label{wsusy}
\eeq
The flow between the two fixed points is generated by another solution,
$\wir$, not supersymmetric and not analytic at $\phi=0$ that can only
be found numerically. The two functions are plotted for comparison
in figure~\ref{figagainst}. 
The function $\wir$ is rather tricky to find directly from (\ref{sp})
but it can be constructed a posteriori once the solution to the
Lagrangian system (\ref{lagrange}), (\ref{lagrange2}), (\ref{constraint}) 
has been found numerically. Such a solution for $\phiir$ is presented in
figure~\ref{figconnect} and can be easily seen to interpolate
between the UV and IR fixed points.
In fact, once the solution $\phiir$ is found, $\wir$ can be defined as
\beqs
    \wir(z) &=& \int_{-\log 2 /\sqrt{5}}^z dw \;
             \dot\phiir\bigg(\phiir^{-1}(w)\bigg)
              - \frac{5}{2^{1/5}\sqrt{3}} \nonumber \\ &=& 
           \int_{-\infty}^{\phiir^{-1}(z)} dy \;\dot\phiir(y)^2
            - \frac{5}{2^{1/5}\sqrt{3}}~.
       \label{construct}
\eeqs
The constant in (\ref{construct}) is chosen to agree with (\ref{sp})
at the IR point. 
It is interesting to analyze the behaviors of $\ws$ and $\wir$ near
the origin. Obviously, $\ws$ is analytic and
\beq 
     \ws(0)=-\frac{5}{2},\quad \ws^\prime(0)=0, 
     \quad \ws^{\prime\prime}(0) = -2, 
     \quad \ws^{\prime\prime\prime}(0) =\frac{6}{\sqrt{5}}, 
\eeq
whereas $\wir$ is not analytic, since
\beq 
     \wir(0)=-\frac{5}{2},\quad \wir^\prime(0)=0, 
     \quad \wir^{\prime\prime}(0) = -1, 
     \quad \wir^{\prime\prime\prime}(0) = \infty. \label{nonana}
\eeq
The two solutions $\ws$ and $\wir$ act as boundaries for a continuum
set of solutions that lay between them, all of which have the same
behavior as (\ref{nonana})
\footnote{There are also solutions with only 
an extremum at the IR point which we do not consider~\cite{sugra7new}.}.

Since the second derivative of $W$ determines whether the behavior of
$\phi$ at $y\to +\infty$ is square-integrable or not, we see that $\wir$
gives rise to a non-square-integrable behavior, thus 
corresponding to deforming the fixed point Lagrangian by $\op$.

\begin{figure}
\begin{center}
\epsfig{file=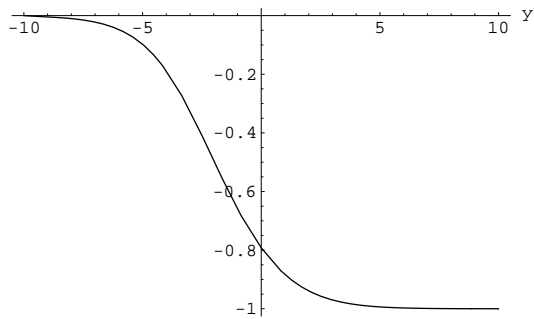,width=7 cm}
\caption{The asymptotic behavior of $\dot\phi/\phi$ shows that 
$\phi\approx e^{-y}$ when $W=\wir$ corresponding to a true deformation
by $\op$.}
\label{figtest}
\end{center}
\end{figure}

In fact {\it none} of the other solutions (including the
superpotential) is physically acceptable in the region $\phi<0$ because
they would correspond to giving a negative vev to $\op$, a manifestly
positive operator. 
If we write the asymptotics of $\phi$ as\footnote{For simplicity we
do not write the polynomial corrections. Also recall that $r^2=-15/V(0)=4$}
\beq
    \phi\approx A e^{-2y/r} + B e^{-4y/r}
\eeq
the above analysis shows that $A=0$ for $\ws$ and non-zero for the
others. This is shown in figure~\ref{figtest} for the particularly
interesting case where the generating function is $\wir$.
If we take the case of $\ws$, so that the vev becomes the
leading term, we get $B<0$, since we are studying the region
$\phi<0$. The term $B$ should still remain negative by continuity as we use
generating functions laying between $\ws$ and $\wir$ and it
will reach zero at $\wir$, precisely as $A$ reaches zero at the
opposite end ($\ws$). $B$ corresponds to a vev for $\op$ and therefore
a negative value must be excluded. It would be nice to check numerically that
our picture is correct but it is rather difficult to isolate the
sub-leading term when $A\not=0$.

\begin{figure}
\begin{center}
\epsfig{file=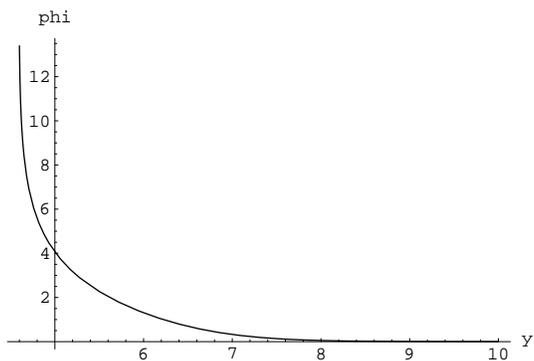,width=7 cm}
\caption{The supersymmetric runaway solution $\phis$
corresponding to a non-conformal vacuum where $\langle\op\rangle > 0$,
plotted against the scale factor $y$ in the vicinity of the UV fixed point.}
\label{figrun}
\end{center}
\end{figure}

Since the leading exponential behavior for all generating functions
is known explicitly, we can be more precise and analyze
the differences between positive and negative $\phi$. From (\ref{hj1})
and (\ref{hj2}), following~\cite{gubser} and 
doing the asymptotics for large $|\phi|$, one can
easily see that the metric has the following behavior, (shifting the 
singularity to $y=0$)
\beqs
  \phi>0:\quad ds^2 &=& y^2 \; \eta_{\mu\nu}dx^\mu dx^\nu +dy^2\label{good}\\
  \phi<0:\quad ds^2 &=& y^{1/8}\; \eta_{\mu\nu}dx^\mu dx^\nu +dy^2\label{bad}.
\eeqs
Solution (\ref{bad}) corresponds to a naked time-like singularity and
our analysis says that it should be excluded. On the other hand, the
runaway solution for $\phi>0$ is acceptable and plotted in
 the neighborhood of the UV point in figure~\ref{figrun}.
This solution corresponds to going to new non-supersymmetric vacua
where $\langle\op\rangle >0$, which for the special case of the superpotential
corresponds to a supersymmetric vacuum. The reason we have many acceptable 
generating functions for $\phi>0$ is that they simply correspond to 
different vev's (or vacua), in contrast to the $\phi<0$ case.

\section{Squashing deformations of $d=4$ supergravity}

The final issue we discuss is the behavior of
${\cal N}=1$, $d=3$ superconformal field theories obtained from
solutions of $d=11$ supergravity~\cite{orig} on squashed
seven-manifolds. These examples are of 
interest because they involve fields from higher levels of 
the Kaluza-Klein tower. They were 
recently considered in~\cite{ar} -- we would like to make a few 
remarks complementing that analysis. 

The two main examples of manifolds allowing for a squashed solution are 
the seven-sphere and the manifold $N(1,1)$~\cite{pagepope}.
Squashed metrics are Einstein metrics, obtained by 
stretching the original one in some directions. 
$N(1,1)$ is a particular instance of a class of seven-dimensional Einstein
manifolds named $N(p,q)$~\cite{pagepope}, which has the peculiarity of
preserving ${\cal N}=3$ supersymmetries, whereas its squashed version
has ${\cal N}=0$.\footnote{However, the orientation-reversed or 
``skew-whiffed'' solution is supersymmetric, with 
${\cal N}=1$~\cite{sugra11}.}
By the AdS/CFT correspondence all these solutions ought to correspond to some
conformal limits of three-dimensional QFT representing the degrees of 
freedom living on M2-branes placed at the apex of the cone over the
compactification manifolds~\cite{conical}.

By the standard procedure, one reads off the global symmetries (``flavor'' and
R-symmetry) from the isometries of the corresponding solution. Once an 
appropriate guess for the gauge group is made, 
it is possible to get a detailed mapping between
the operators and the fields of the KK spectrum, based on matching
supersymmetry representations. The theory dual to $S^7$ is
a 3 dimensional ${\cal N}=8$ CFT with $SO(8)$ R-symmetry group, while
$N(1,1)$ gives rise to a CFT with $SU(3)\times SU(2)$ global
symmetries and supercharges transforming in the {\bf 3} of $SU(2)$. 
Moreover, the
situation is more difficult for squashed solutions. Having ${\cal N}=1$
in three dimensions there is no R-symmetry and the usual procedure  
does not apply straightforwardly. 
The case of $N(1,1)$ is particularly puzzling as
squashed and unsquashed solutions share the same global symmetries. 

To study the possibility of having
domain-wall solutions interpolating between such theories
one considers a truncation of the Kaluza-Klein spectrum and derives
an effective four-dimensional action for the non-zero fields.  
The potential 
for the sphere is known from the work of~\cite{squashed} in terms of two 
scalars $u$ and $v$ appearing in the eleven dimensional metric as
\beq
      ds^2 =  e^{-7u} ds^2(AdS_4) + e^{2u+3v} ds^2(\mathrm{base}) 
      +e^{2u-4v} ds^2(\mathrm{fibre})~, \label{metric}
\eeq
where the seven-sphere is thought of as a $S^3$ fibration over the base 
$S^4$.
The potential for the squashed $N(1,1)$ has been given in terms of four 
scalars in~\cite{fourscalars} but for our purposes it is sufficient to 
repeat the computation of~\cite{squashed} using (\ref{metric})
where now the base manifold is $CP^2$ and the fiber is $RP^3$, thus 
obtaining a potential also dependent only on two scalars. In both cases the 
potential can be written as
\beq
       V(u,v) = \lambda e^{-9u} \left(\alpha e^{4v} - e^{3v} -
       \frac{1}{32\alpha} e^{-10 v}\right) + 2Q^2 e^{-21 u}, 
       \label{vuv}
\eeq
where $Q$ is the Page charge and, for the sphere, $\alpha=-1/8$ and 
$\lambda=48$, whereas,
for $N(1,1)$, we have $\alpha=-1/16$ and $\lambda=24$. Amusingly, all the 
physical quantities, such as the conformal dimensions for the operators, 
turn out to be independent of $\alpha$ and $\lambda$.

The potential (\ref{vuv}) has two fixed points but the field $u$ always 
describes a non-renormalizable (irrelevant) operator. From the equivalent 
mechanical 
problem, the flow between these two points would have to connect a maximum 
of $-V$ to a saddle point of $-V$, clearly an unstable situation -- 
contrary to the situation occurring for some flows in $d=5$ gauged 
supergravity, where the ``particle'' rolls along a valley from a saddle point 
to a minimum. 

As mentioned in the introduction, if the RG
equations where truly first order, one could argue from general
theorems that there must still be a critical line connecting the
points. However, the equations expressed in terms of the potential are
second order and the existence of this solution is not guaranteed in this
case. After some numerical tests we now believe that there is no
such flow.

The potential (\ref{vuv}) has another peculiar property: It is possible 
to find explicitly one generating function $W$ that has one critical point 
at the squashed solution. The function is\footnote{We take the case of the
sphere for concreteness. For the $N(1,1)$ case the factor 3 in front of
$e^{2v}$ is changed to 3/2.}
\beq
   W(u,v)=-\frac{1}{\sqrt{8}}e^{-\frac{9}{2}u}\Big(3e^{2v}+6e^{-5v}-
|Q|e^{-6u}\Big), \label{supsup}
\eeq
which is a solution to
\beq
   V=\frac{16}{63}(\partial_{u}W)^{2}+\frac{8}{21}(\partial_{v}W)^{2}-12W^{2}.
\eeq
At first, it seems 
rather counterintuitive that the point with {\it less} supersymmetries 
should appear as an extremum, 
but we must remember that we are not dealing with a gauged
supergravity, where only low-lying KK excitations are included.
Still, it is tempting to believe that 
the solution associated with $W$ describes different {\it supersymmetric}
vacua of the theory. 
As a check one can show, expanding $W$ near its critical point, 
that the solution corresponds to the operator associated to $v$ 
getting a vev -- more specifically $v\sim \exp(- 5y/3r)$.
There is a choice between a theory in which the conformal dimension 
of the operator is $\Delta = 5/3$ or $\Delta = 4/3$,
which is also allowed~\cite{plusminus}. 
Finally, one finds that the runaway solution also satisfies the 
criterion~\cite{gubser} of boundness 
from above of the potential. Hopefully, further investigations will 
reveal if even these models consistently describe holographic RG flows.

\section{Acknowledgments}
We wish thank U. Boscain, U. Gran, J. Kalkkinen and A. Tomasiello for useful
discussions. D.M. wishes to thank Chalmers University of G\"oteborg for 
financial support and
kind hospitality when this work was initiated,
and acknowledges partial support from EU TMR program CT960045.


\begin{thebibliography}{99}

\bibitem{thethree}
J.~Maldacena, 
''The Large N limit of superconformal field theories and supergravity,''
Adv. Theor. Math. Phys. {\bf 2}, 231 (1997)
hep-th/9711200. \hfill\break
S.S.~Gubser, I.R.~Klebanov and A.M.~Polyakov,
``Gauge theory correlators from non-critical string theory,"
Phys. Lett. {\bf B428}, 105 (1998)
hep-th/9802109. \hfill\break
E.~Witten,
``Anti-de Sitter space and holography,"
Adv. Theor. Math. Phys. {\bf 2}, 253 (1998)
hep-th/9802150.\hfill\break

\bibitem{rg}
D.~Z.~Freedman, S.~S.~Gubser, K.~Pilch and N.~P.~Warner,
``Renormalization group flows from holography supersymmetry 
and a  c-theorem,'' hep-th/9904017.\hfill\break
J.~Distler and F.~Zamora,
``Non-supersymmetric conformal field theories from stable 
anti-de Sitter  spaces,''
Adv.\ Theor.\ Math.\ Phys.\  {\bf 2}, 1405 (1999)
hep-th/9810206.\hfill\break

\bibitem{gira}
L.~Girardello, M.~Petrini, M.~Porrati and A.~Zaffaroni,
``Novel local CFT and exact results on perturbations of N = 4 super 
Yang-Mills from AdS dynamics,'' JHEP {\bf 9812}, 022 (1998)
hep-th/9810126.\hfill\break
A.~Khavaev, K.~Pilch and N.~P.~Warner,
``New vacua of gauged N = 8 supergravity in five dimensions,''
hep-th/9812035.\hfill\break

\bibitem{plusminus}
V.~Balasubramanian, P.~Kraus and A.~Lawrence,
``Bulk vs. boundary dynamics in anti-de Sitter spacetime,''
Phys.\ Rev.\  {\bf D59}, 046003 (1999)
hep-th/9805171.\hfill\break
V.~Balasubramanian, P.~Kraus, A.~Lawrence and S.~P.~Trivedi,
``Holographic probes of anti-de Sitter space-times,''
Phys.\ Rev.\  {\bf D59}, 104021 (1999)
hep-th/9808017.\hfill\break
I.~R.~Klebanov and E.~Witten,
``AdS/CFT correspondence and symmetry breaking,''
Nucl.\ Phys.\  {\bf B556}, 89 (1999)
hep-th/9905104.

\bibitem{bound}
L.F.~Abbott and S.~Deser,
``Stability Of Gravity With A Cosmological Constant,"
Nucl. Phys. {\bf B195}, 76 (1982).\hfill\break
P.~Breitenlohner and D.Z.~Freedman,
``Stability In Gauged Extended Supergravity,"
Ann. Phys. {\bf 144}, 249 (1982). \hfill\break
L.~Mezincescu and P.K.~Townsend,
``Stability At A Local Maximum In Higher Dimensional Anti-De Sitter Space 
And Applications To Supergravity,"
Ann. Phys. {\bf 160}, 406 (1985). \hfill\break
L.~Mezincescu, P.K.~Townsend and P.~van Nieuwenhuizen,
``Stability Of Gauged D = 7 Supergravity And The Definition Of 
Masslessness In (Ads) In Seven-Dimensions,"
Phys. Lett. {\bf 143B}, 384 (1984). \hfill\break

\bibitem{vevdef}
L.~Girardello, M.~Petrini, M.~Porrati and A.~Zaffaroni,
``The supergravity dual of N = 1 super Yang-Mills theory,''
hep-th/9909047.

\bibitem{hj}
J.~de Boer, E.~Verlinde and H.~Verlinde,
``On the holographic renormalization group,''
hep-th/9912012.\hfill\break
E.~Verlinde and H.~Verlinde,
``RG-flow, gravity and the cosmological constant,''
hep-th/9912018.

\bibitem{dolan}
B.~P.~Dolan,
``Symplectic geometry and Hamiltonian flow of 
the renormalization group equation,''
Int.\ J.\ Mod.\ Phys.\  {\bf A10}, 2703 (1995)
hep-th/9406061.

\bibitem{sugra7}
M.~Pernici, K.~Pilch and P.~van Nieuwenhuizen,
``Gauged Maximally Extended Supergravity In Seven-Dimensions,''
Phys.\ Lett.\  {\bf B143}, 103 (1984).\hfill\break
L.~Mezincescu, P.~K.~Townsend and P.~van Nieuwenhuizen,
``Stability Of Gauged D = 7 Supergravity And The Definition Of 
Masslessness In (Ads) In Seven-Dimensions,''
Phys.\ Lett.\  {\bf B143}, 384 (1984).\hfill\break
P.~K.~Townsend and P.~van Nieuwenhuizen,
``Gauged Seven-Dimensional Supergravity,''
Phys.\ Lett.\  {\bf B125}, 41 (1983).

\bibitem{sugra7new}
K.~Skenderis and P.~K.~Townsend,
``Gravitational stability and
renormalization group flow,''
Phys.\ Lett.\  {\bf B468}, 46 (1999)
hep-th/9909070.

\bibitem{multiplet1}
M.~G\"unaydin, P.~van Nieuwenhuizen and N.~P.~Warner
``General construction of the unitary representations of anti-de Sitter 
superalgebras and the spectrum of the $S^4$ compactification of 
11-dimensional supergravity'',
Nucl.\ Phys.\  {\bf B255}, 63 (1985).

\bibitem{multiplet2}
P.~van Nieuwenhuizen
``The complete mass spectrum od $d=11$ supergravity compactified on $S_4$ 
and a general mass formula for arbitrary cosets $M_4$'',
Class. Quantum Grav. 1 (1985).

\bibitem{gubser}
S.~S.~Gubser,
``Curvature singularities: The good, the bad, and the naked,''
hep-th/0002160.

\bibitem{sugra11}
For a review of the extensive literature on the subject, see\hfill\break
M.~J.~Duff, B.~E.~Nilsson and C.~N.~Pope,
``Kaluza-Klein Supergravity,''
Phys.\ Rept.\  {\bf 130}, 1 (1986) and references therein.

\bibitem{ar}
C.~Ahn and S.~Rey,
``Three-dimensional CFTs and RG flow from squashing M2-brane horizon,''
hep-th/9908110.
\hfill\break
C.~Ahn and S.~Rey,
``More CFTs and RG flows from deforming M2/M5-brane horizon,''
hep-th/9911199.

\bibitem{orig}
E.~Cremmer, B.~Julia and J.~Scherk,
``Supergravity theory in 11 dimensions,''
Phys.\ Lett.\  {\bf B76}, 409 (1978).

\bibitem{pagepope}
G.~Jensen, Duke Math. J. {\bf 42}, 397 (1975). \hfill\break
M.~A.~Awada, M.~J.~Duff and C.~N.~Pope,
``N = 8 supergravity breaks down to N = 1,''
Phys.\ Rev.\ Lett.\  {\bf 50}, 294 (1983). \hfill\break
M.~J.~Duff, B.~E.~Nilsson and C.~N.~Pope,
``Spontaneous Supersymmetry Breaking By The Squashed Seven Sphere,''
Phys.\ Rev.\ Lett.\  {\bf 50}, 2043 (1983).\hfill\break
L.~Castellani and L.~J.~Romans,
``N=3 And N=1 Supersymmetry In A New Class Of 
Solutions For D = 11 Supergravity,''
Nucl.\ Phys.\  {\bf B238}, 683 (1984).\hfill\break
D.~N.~Page and C.~N.~Pope,
``New Squashed Solutions Of D = 11 Supergravity,''
Phys.\ Lett.\  {\bf B147}, 55 (1984).

\bibitem{conical}
J.~M.~Figueroa-O'Farrill,
``Near-horizon geometries of supersymmetric branes,''
hep-th/9807149.\hfill\break
B.~S.~Acharya, J.~M.~Figueroa-O'Farrill, C.~M.~Hull and B.~Spence,
``Branes at conical singularities and holography,''
Adv.\ Theor.\ Math.\ Phys.\  {\bf 2} (1999) 1249
hep-th/9808014.\hfill\break
D.~R.~Morrison and M.~R.~Plesser,
Adv.\ Theor.\ Math.\ Phys.\  {\bf 3}, 1 (1999)
hep-th/9810201.

\bibitem{squashed}
D.~N.~Page,
``Classical Stability Of Round And Squashed Seven 
Spheres In Eleven-Dimensional Supergravity,''
Phys.\ Rev.\  {\bf D28}, 2976 (1983).

\bibitem{fourscalars}
O.~Yasuda,
``Stability Of Englert Type Solutions On N(pqr) In D = 11 Supergravity,''
Phys.\ Rev.\  {\bf D31}, 1899 (1985).

\end{thebibliography}
\end{document}